# Strategic decision points in experiments: A predictive Bayesian optional stopping method

Xiaomi Yang [a*], Carol Flannagan [a,b], Jonas Bärgman [a]

[a] Chalmers University of Technology, Vehicle Safety Division at the Department of Mechanics and Maritime Sciences, Chalmersplatsen 4, 412 96 Göteborg, Sweden

[b] University of Michigan Transportation Research Institute, 2901 Baxter Rd, Ann Arbor, MI 48109, United States

∗ Corresponding author. Tel.: +46-772 6901; E-mail: xiaomi.yang@chalmers.se

## Abstract

Sample size determination is a critical element in experimental design – not least in traffic and transport research. On one hand, without a sufficient sample size, an experiment cannot reliably answer its research question(s); on the other hand, too many samples are a waste of resources. Frequentist statistics require that the sample size be determined up front and the experiment be carried through to completion. That approach relies on a power analysis and its assumptions, which cannot be adjusted once the experiment has started. Bayesian sample size determination, with proper priors, can replace this approach. Bayesian optional stopping (BOS) iteratively analyzes collected data and enables stopping the experiments if the analyzed results show the statistical target (precision or significance) has been reached. We combined BOS with Bayesian rehearsal simulations in a method we call predictive Bayesian optional stopping (pBOS). Bayesian rehearsal simulations generate possible future data based on the Bayesian posterior distribution of some initially collected data and selected priors. Like BOS, pBOS iteratively analyzes the current experiment results and stops the experiment when a pre-defined statistical target is reached. Moreover, the rehearsal simulations also iteratively predict how the experiment might unfold, so that it can be stopped if the target is unlikely to be reached with some fixed maximum number of samples (typically constrained by available resources). While developing the method, we identified an inherent bias in the predictions. To correct for it, we employed a multiple linear regression model on the actual precision and predicted precision of the targets. We demonstrate the benefits and drawbacks of pBOS compared to BOS and frequentist power analysis, as well as providing guidelines for its use. The pBOS method shows a cost benefit up to 118% better than the cost benefit of the traditional BOS, depending on characteristics of the experiment, which in turn is better than the frequentist sample size determination. In summary, a researcher using pBOS can stop an experiment when the research question cannot be answered with the allocated resources, or when sufficient data have been collected—providing potential redirection of resources or cost-savings compared with traditional frequentist sample size determination or BOS.

# 1 Introduction
## 1.1 Empirical experiments and cost constraints
Empirical experiments are widely used to address a wide range of research questions in the domain of traffic safety (D. Albers et al., 2020; Butakov & Ioannou, 2015; Ettema et al., 2010; Goodin et al., 2019; Jiang et al., 2018; T. Li et al., 2021; Y. Li et al., 2020; Sieber & Farber, 2016). Unlike observational studies, experiments enable manipulation of one or several independent variables while minimizing the impact of confounding factors. However, experiments are often expensive (Lakens, 2022). Further, a study that is too small may not yield valuable results, leading to wasted resources, while a study that is too large may consume more resources than required. Resource constraints often limit how extensive a study can be (Lakens, 2022; Lenth, 2001). Therefore, determining the number of samples needed in a specific study has been a crucial challenge in experiment planning for many years (Adcock, 1997; Balki et al., 2019; Rahman, 2023).

## 1.2 Sample size determination

The approaches for determining the required sample size of an experiment can be divided into two groups: a) those that estimate the needed sample size "up front" before data have been collected (or after some pilot data have been collected), and b) those that stop the experiment when "enough" samples have been collected (this is also called optional stopping; Rouder, 2016). The former approach can help estimate whether a study can be completed with the available resources, and the latter can help minimize the cost of an experiment; after enough data are collected, collecting more is a waste of resources.

### 1.2.1 Traditional frequentist sample size determination

The most common up front sample size determination approach is frequentist power analysis (Dubin, 1990; Wassertheil & Cohen, 1970). Frequentist power quantifies the probability of correctly rejecting the null hypothesis (Cohen, 1992). Thus a frequentist power analysis can be used to estimate the sample size needed to obtain a specific probability of detecting a particular (true) result, given constraints on the probability of falsely rejecting the null hypothesis. This estimation allows the experimenter to control the probabilities of Type I and Type II errors when rejecting the null hypothesis (Inoue et al., 2012). (A Type I error is concluding there is an effect when there is not, while a Type II error is failing to detect an effect when there is one). The goal of sample size determination is to ensure the desired probabilities of these two errors with the smallest possible sample size, based on assumptions about the true underlying variance in the dependent variable and the effect size. The variance of the dependent variable indicates how much sample values differ from the mean on average; effect size is the magnitude of the difference or relationship being studied. Frequentist power analysis is typically effect size-based.

An alternative is precision-based sample size determination, which aims to achieve a specific level of precision by evaluating the variance of the parameter of interest. (Precision is the inverse of variance.) The target for frequentist precision-based sample size determination is commonly a desired confidence interval width. The term "target" here means the desired value of the specific statistic in question; a confidence interval is a range of parameter values that contains the true population parameter with a given probability or "confidence level", typically 95%. In summary, the effect size-based target focuses on the sample mean while the precision-based target focuses on the sample variance.

Frequentist power analysis gives an indication of the resources needed for a given experiment, but the assumptions about effect size and variance of the dependent variable might not be accurate (C. Albers & Lakens, 2018; Leon et al., 2011). The frequentist hypothesis testing paradigm has an additional limitation: once a study has started, it must be run to completion (Fried & Peterson, 1969). Resource-wise, it would be better if there were a way to stop the experiment when enough data have been collected. The issue is that "peeking at the data" (Kruschke, 2014, p. 300) in frequentist statistics leads to an increased probability of type I errors and introduces a bias in the data interpretation. Examples of this effect have been provided by Yu et al. (2014) and Kruschke (2014).

The choice of target statistic (e.g., effect size or precision) and the target value depend on the research question and the experiment. Effect size-based sample size determination focuses on quantifying the magnitude of the effect being investigated, such as the differences between groups. In contrast, precision-based sample size determination focuses on achieving a specific standard error for the parameter of interest. For a detailed discussion of the difference between these types of targets, see Kelter (2023).

#### 1.2.1.1 Bayesian alternatives

Unlike frequentist statistics, Bayesian statistics allows an experiment to be stopped partway through based on the results of data collected during the experiment—without violating underlying assumptions of the method (Rouder, 2014). The aim is to stop sequential experiments early when enough data have been collected to reach some statistical target in the Bayesian statistics framework. In this work, this process is called Bayesian optional stopping, or BOS (Edwards et al., 1963; Good, 1991; Lindley, 1957). Other terms have been used, such as *early stopping* (Flournoy & Tarima, 2023; Tarima & Flournoy, 2022), *Bayesian sequential decision-making* (Wan et al., 2023), *sequential Bayes factors test* (Pourmohamad & Wang, 2023; Schönbrodt et al., 2017; Stefan et al., 2022), and *Bayesian stopping* (Douven, 2023). Note that BOS is iterative and stops only based on the data that are already collected and the prior used.

#### 1.2.1.2 BOS stopping criteria

Stopping-criteria selection is a crucial part of BOS, as stopping criteria define under what target condition the experiment can be stopped. Stopping criteria are established based on target statistics. There are two general types of targets (and thus two types of stopping criteria: effect-size-based and precision-based), and the choice depends on the research question. If the target is reached the research question can be answered (with some confidence). There is ongoing scientific debate about the appropriateness of and issues with the use of effect-testing stopping targets (de Heide & Grünwald, 2021), while there is a consensus that precision-based targets can be used as stopping criteria in BOS. For this work we therefore chose a precision target-based stopping criterion; not only do we avoid that debate, but this type is relatively straightforward method-wise.

#### 1.2.1.3 Bayesian posterior-based sample size determination

Kruschke (2014) proposed going beyond the traditional BOS, using early samples to estimate the number of additional samples needed to reach the target. That is, Kruschke suggested using the Bayesian model's posterior distribution, which combines prior information and collected data, to estimate the probability of achieving the desired statistical target. The estimation involves simulating potential future datasets multiple times, using the posterior distribution as the underlying data-generation model parameter distribution. With a small initial dataset, the results of such simulations—which Kruschke calls "rehearsal simulations"—form the basis for estimating the sample size needed to reach a pre-defined power. Kruschke's method looks into the future, but only once.

#### 1.2.1.4 Predictive Bayesian optional stopping (pBOS)

This work extends existing Bayesian optional stopping approaches by iteratively integrating Krushcke's Bayesian sample size determination approach with newly collected data. The combined approach iteratively looks into the future to estimate the probability of reaching a target within the maximum sample size (often determined by the available resources) during data collection.

### 1.3 The aim, challenges, and utility of pBOS

The main aim of this paper is to develop and evaluate the pBOS method, which provides the opportunity to stop an experiment either when the desired statistical target is reached or when it is estimated that, given the available resources for the experiment there will not be enough data at the end of the experiment to be able to reach the target.

We identified two methodological challenges. First, under most choices of prior, the rehearsal simulations underestimate the precision of the parameter of interest. To address this challenge, we incorporated a calibration step into the method. Second, we could not find a metric in the literature that let us evaluate and compare the cost benefits of various sample-size determination methods that reflects the common issue of limited-resource. We therefore propose such a cost-benefit metric.

In this work we first demonstrate and assess pBOS, using a standard normal distribution N(0,1) as the data model. We then demonstrate the utility of pBOS in the traffic safety domain by applying the method to an experiment which assessed the reaction times of drivers to a forward collision warning (FCW) safety system under development. We further discuss why pBOS may be particularly relevant in the domain of traffic safety. To optimize the utilization of pBOS, we have made the code open-source.

### 1.4 Motivating traffic safety example

In this work we use the assessment of a FCW safety system as an example. Globally around 1.2 million people die in traffic crashes every year, and many more are injured (World Health Organization, 2023). The study of human behavior through empirical experiments is often an integral part of the development of advanced vehicle safety systems that help avoid crashes. The aims of such studies range from investigating driver reactions to some event or system (Green, 2000) to drivers' self-reporting of system acceptance and trust (He et al., 2022). However, experiments with drivers as participants are often expensive to prepare and execute, and often require access to specialized, costly facilities such as driving simulators or test tracks. Because costs are proportional to the duration of driver behavior studies, a method such as BOS or pBOS that can optimize sample size determination and possibly shorten the studies can lead to substantial savings. The resulting funds could be made available for additional development or reducing end-user product costs.

FCW safety systems aim to avoid—or at least mitigate—crash severity by warning drivers of an impending collision so that they can initiate braking or steering (Jamson et al., 2008). The *trigger time,* the time when the FCW warning is provided to the driver, is crucial for determining the FCW's safety performance, but it also affects the driver's acceptance of the system. Earlier trigger times will avoid more crashes, but may produce nuisance warnings; as a result, an annoyed driver might, for example, turn off the system (Yang et al., 2024). Therefore, studying how drivers react to different FCW algorithms is important in order to develop an effective FCW system with high acceptance.

We use data from an FCW experiment that have already been collected to generate more, hypothetical data. Instead of using BOS or pBOS in the actual experiment, we applied the methods to a parametric data-generation model based on the real-world data as if we were actually conducting the experiment (one sample at a time). The implementation and results are presented in Section 4.

## 2 Methods

This section provides a high-level overview of frequentist sample size determination and three key methods: 1) BOS, 2) Kruschke's (2014) rehearsal-simulation-based approach to Bayesian sample size determination, and 3) pBOS. Section 3 (Statistical simulations) and Section 4 (Application in the traffic safety domain) describe in more detail how the methods were applied, assessed, and compared using a precision target through simulations.

### 2.1 Frequentist sample-size determination method

As the investigated stopping target is a precision-based target, the frequentist sample size determination method that we compare pBOS to in this paper is shown in Equation (1) for precision-based target. It calculates the sample size n needed to reach the confidence interval E. Z is the value from the table of probabilities of the standard normal distribution for the desired confidence interval (e.g., Z=1.96 for a 95% confidence interval), and $\sigma$ is the standard deviation of the parameter of interest.

$$n = \left(\frac{Z\sigma}{E}\right)^2 \tag{1}$$

## 2.2 Existing Bayesian methods for sample size determination
### 2.2.1 Traditional Bayesian optional stopping (BOS)

Figure 1 illustrates the BOS sequential decision-making process for determining sample size. An assessment performed after each new sample is drawn determines whether a pre-specified stopping criterion has been reached. One possible BOS stopping criterion is a threshold for the credible interval length (CIL), which is the width of the highest density interval (HDI). The HDI is an interval within a posterior distribution where all parameter values have a higher probability density than those outside the interval (Turkkan & Pham-gia, 1993). HDI is (somewhat) analogous to the frequentist confidence interval, representing the smallest interval that captures a specified percentage of the parameter values. The most commonly used percentage is 95%; its length is called then called 95% CIL.

The decision whether to stop the experiment is made iteratively as new data are collected. The more data collected, the narrower the posterior distribution, which will lead to less parameter variability and, consequently, higher precision in the parameter estimate. In precision-based BOS, the CIL ($T_i$ in Figure 1, whose units are those of the parameter) is the stopping criterion; the target of the experiment is to get the CIL for a certain coverage below the pre-set threshold ($CIL_{thres}$). If the current 95% CIL is larger than the pre-set threshold, the experiment continues, and more data are collected. One can choose to collect one or more new samples ($j$ in Figure 1) before the next CIL threshold check. As new data are collected, a new posterior distribution is generated, and the target is iteratively checked until either the collected data have met the target or the experiment has run out of available resources; then the experiment can stop. The available resources are described by $N_{max}$, the maximum number of samples that can be allocated to the experiment.

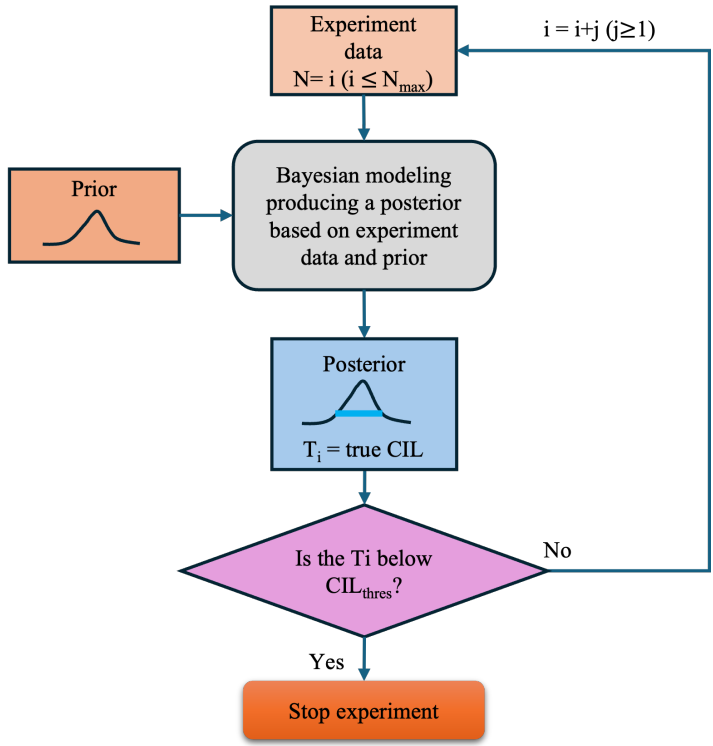

*Figure 1    Illustration of the traditional BOS process. In this figure, i is the sample size of all collected data and j is the sample size of newly collected data. $T_i$ is the collected data CIL and $CIL_{thres}$ is a pre-set threshold for target CIL.*

### 2.2.2    Bayesian posterior-based sample size determination

Bayesian posterior-based sample size determination, as laid out by Kruschke (2014), uses a smaller amount of data (e.g., from a pilot study or from early samples) together with a prior to generate a posterior. The posterior is then assumed to represent the "true" data-generation model. Given this posterior distribution and the chosen priors, rehearsal simulations produce *m* sets of future data (Figure 2). A distribution of the CILs is created based on the *m* groups of posterior distributions (from the *m* sets of data). Finally, the maximum tolerated CIL (TL's CIL in Figure 2) is chosen by taking the tolerance level (TL) percentile of the CILs' distribution, where TL is the selected target threshold probability of achieving the desired CIL after $N_{max}$ trials. If that maximum tolerated CIL value is smaller than the precision-based target ($CIL_{thres}$), it is likely (with the probability of TL) that the precision target will be reached if *k* future samples are collected. With this approach it is possible to estimate how many data samples are likely to be needed to reach $CIL_{thres}$ with a specific TL.

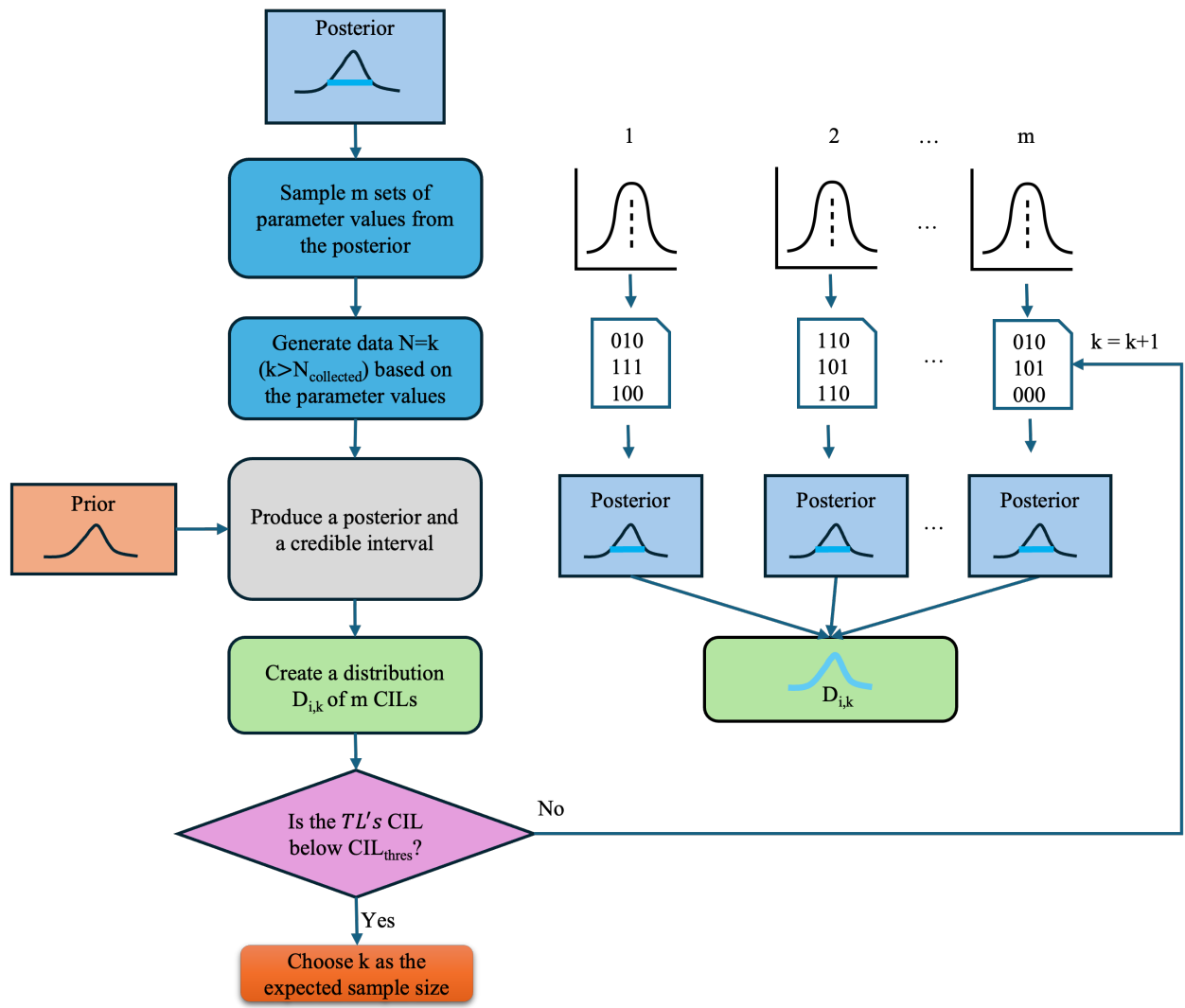

*Figure 2    Illustration of the rehearsal simulations. The tolerance level (TL), the estimated probability of reaching the target, is the term used here to avoid confusing it with the frequentist term "power" (used by Kruschke; 2014). In addition to the notations in Figure 1, $N_{collected}$ is the sample size of the collected data. k is the sample size for predicted future data and m describes how many groups of parallel experiments for possible future data collection. $D_{i,k}$ is a distribution of CIL.*

## 2.3   Predictive Bayesian optional stopping (pBOS)

The pBOS method extends the traditional BOS framework by performing the rehearsal simulations iteratively. Studies based on pBOS can benefit from the updating of both the current status of the posterior (relative to the desired statistical target) and the expected number of samples needed to reach that target (if it has not yet been reached). For studies that include resource-intensive data gathering, it can be as useful to know when to stop an experiment and try something else as to know when the statistical target has been achieved.

In the process of developing pBOS, we noted that—with the exception of central informative priors as introduced in Section 3.1.1.4 — the rehearsal simulations consistently overestimated CILs relative to the ground truth. To compensate, we proposed an additional regression-based calibration step. The following sections describe the basics of the pBOS and the calibration process.

### 2.3.1 pBOS without calibration

Figure 3 illustrates the pBOS process. Part A is a basic Bayesian model; during pBOS initialization it combines the initial experiment data with the prior and outputs the posterior distribution. This first step is the same as the first step in BOS—collecting a small set of initial data and merging it with a prior using Bayes rule, producing a posterior distribution. If the posterior suggests that the stopping threshold ($CIL_{thres}$) has been reached already (which is unlikely during initialization, when there is only a relatively small amount of data), then the experiment can be stopped, using the same rule as in BOS. Otherwise, more rehearsal simulations based on this initial posterior distribution are conducted to generate possible future data, up to a maximum sample size ($N_{max}$).

As shown in part B in Figure 3, a posterior distribution from the simulated-future data sample is calculated by combining the data from each rehearsal simulation with the prior. As m groups of rehearsal simulations are performed, m groups of possible data are generated, which naturally result in m posterior distributions. Each simulated data posterior distribution has a CIL. This results in a CIL distribution, with m CIL values and a median CIL. The median CIL is used in the calibration process (see Part C and Figure 4). The output of the calibration (described in further detail in Section 2.3.1.1) is the size of the offset between the median of the CIL distribution for the maximum sample size ($N_{max}$) and the true $CIL_{Nmax}$. This offset is used to recalibrate the CIL distribution, which is then used to decide whether to stop the experiment (by comparing the shifted median CIL to the $CIL_{thres}$).

To create the stopping point, in addition to the target threshold, it is necessary to predefine a TL. The TL is the probability of reaching the statistical target (here CIL). In practice, the TL defines the percentile value from the CIL distribution that is to be compared with the CIL target (pre-defined $CIL_{thres}$). If the TL-specific CIL is smaller than the $CIL_{thres}$, the experiments should continue, as the simulated-future samples suggest that it should be possible to reach the CIL target with at least the pre-defined TL probability if $N_{max}$ data are collected. However, if the (calibrated) TL-specific CIL is larger than $CIL_{thres}$, the experiment could be stopped, as it is not likely (with the given tolerance probability) that the target will ever be reached.

If the experiment continues, new data is collected (shown as i =i+j in Part A of Figure 3). Then, for the next iteration of pBOS, the traditional BOS assessment, including the new data, is applied, followed by the next iteration of the rehearsal simulations and an assessment of the median CIL, as described above. Iterations continue until the target has been reached or the rehearsal simulations indicate that it is unlikely that the target will be reached before $N_{max}$.

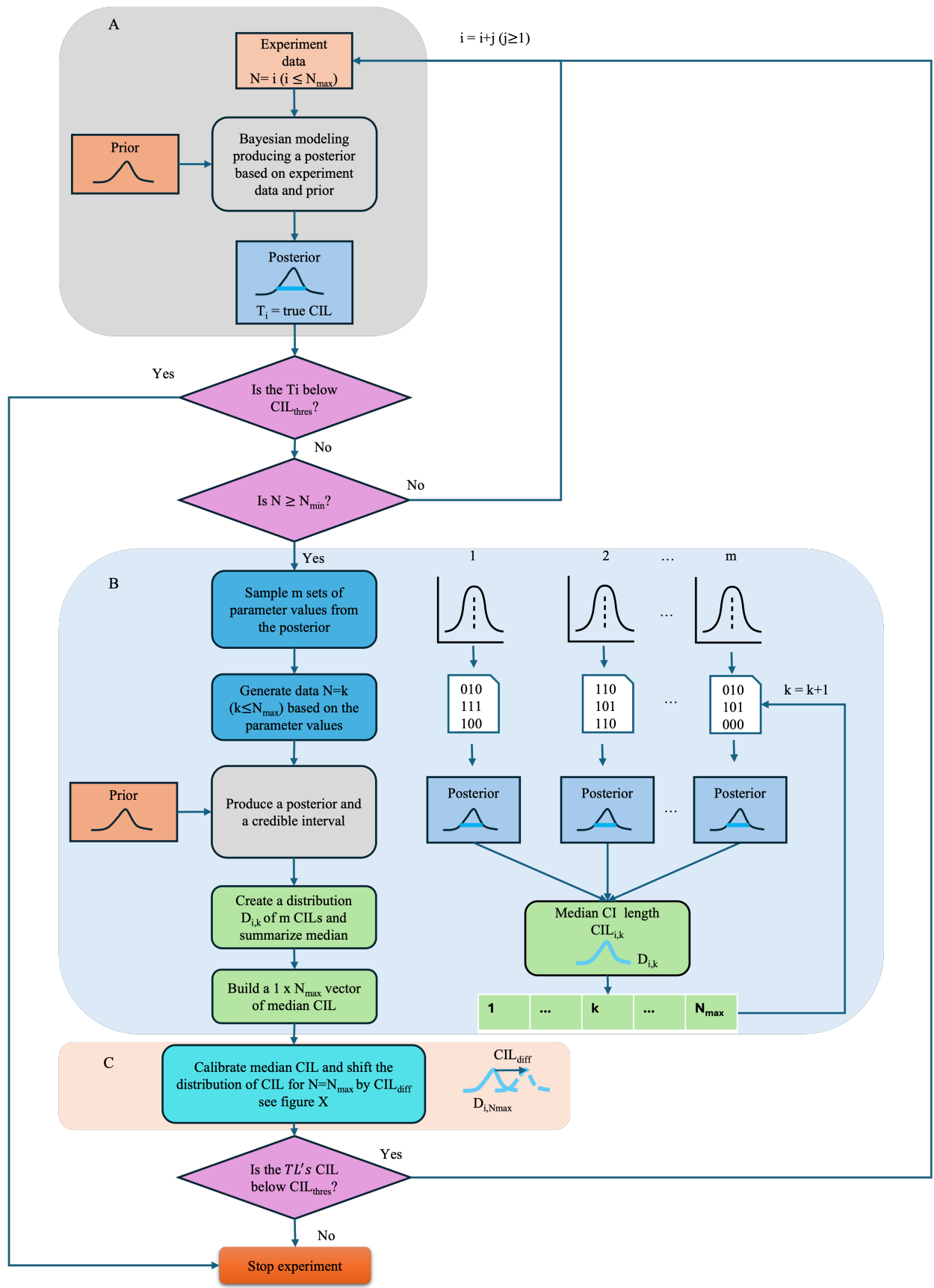

*Figure 3      Illustration of the pBOS process. Part A is a basic Bayesian model. Part B illustrates rehearsal simulations, and Part C is the calibration process for the predicted CIL. In addition to notations in Figure 1 and Figure 2, $N_{max}$ is the maximum number of samples that can be allocated to the experiment and $CIL_{diff}$ is the difference between predicted CIL and the $T_i$ true data CIL.*

2.3.1.1   Regression model for calibration

As previously mentioned, during the development and assessment of pBOS, our simulations showed that the approach of using the posterior to predict future samples produced an inherent bias of the precision (CIL) estimate (we later confirmed this analytically; see Appendix A). The following section describes the bias in the precision (CIL) estimate and proposes a method to address it.

As the sample size increases, the probability of reaching the target also increases. In pBOS each of the *m* parameter groups from the posterior distribution (see Figure 3) describes an assumed underlying data generation model for the experiment. For most choices of prior (other than informative priors), this model is biased towards CILs that are larger than those in the true data generation model, because the variance of the true data posterior and the variance of the generated-data posteriors are not identical. Unlike the true data, the variance of the generated posteriors vary across the *m* parameter groups, but the median variance tends (typically) to be larger than the true data variance, resulting in overestimation of CIL and underestimation of the ability to reach the goal (See Appendix A for details).

This larger variance/CIL and concomitant lower precision estimate in the absence of real data result in a more pessimistic decision by pBOS. As a result, without calibration, pBOS might recommend stopping the experiment even when it is likely that the statistical target can be reached. Also note that the more variance there is in the posterior distribution based on the actual available experiment data, the more variance there will be in the data generation model. Thus, the smaller the sample size of the experiment data used to project into the future, the wider the posterior distribution and the bigger the variance in the assumed data generation model. Consequently, the underestimation of the precision is larger early in the pBOS iterations.

To correct for the underestimation of the posterior metric in the pBOS process, we introduced a calibration step, illustrated in Figure 4. We used a regression model because pBOS is iterative, so it is possible to compare the *future data's posterior estimates* from the rehearsal simulations to the *actual data's posterior estimate* after each new sample is collected during the process. As shown in Figure 3, during rehearsal simulations, the median of the distribution of the CILs for a large set of predicted (future) data is used as the CIL estimate. Based on this estimate, for each collected group of data N = i, a vector of median CILs is generated. As shown in Figure 3, the true data CIL can be grouped together with the predicted median CIL into one i X (1+$N_{max}$) table. The first column is the data CIL and the other $N_{max}$ columns are the predicted median CIL. This table can then be used as input to the calibration model. To avoid including high-uncertainty data, only data above the current $N_{min}$ (in Figure 4, $N_{min}$ =4) is used in the regression.

In the regression model, the dependent variable is the collected data's actual CIL. As precision is the inverse of variance, parameters that influence the variance also influence precision. These parameters become the predictors in the regression model. The predictors used in the regression are a) the sample size of collected samples i, and b) the sample size of the predicted data k. The larger the difference between the predicted data sample size and the collected data sample size, the more variance is introduced in the prediction, and the lower the precision estimation (here CIL) is. Consequently, the predicted CIL is dependent on i and k. To fit the regression model, the CIL predictor and dependent variable is modeled as $\frac{1}{T_i^2}$. The inverse square form was chosen based on the analytical expression of variance (see Appendix A for details about the rationale for this choice). The regression model is shown in equation (2), where the $\beta s$ are the coefficients, $i$ the number of data samples, $k$ the number of predicted data, $CIL$ the predicted median CIL, and $T_i$ the true data CIL. The true CIL is the posterior CIL based on i and the prior.

$$\frac{1}{T_i^2} \sim \beta_0 + \beta_1 \frac{1}{CIL^2} + \beta_1 i + \beta_2 k \tag{2}$$

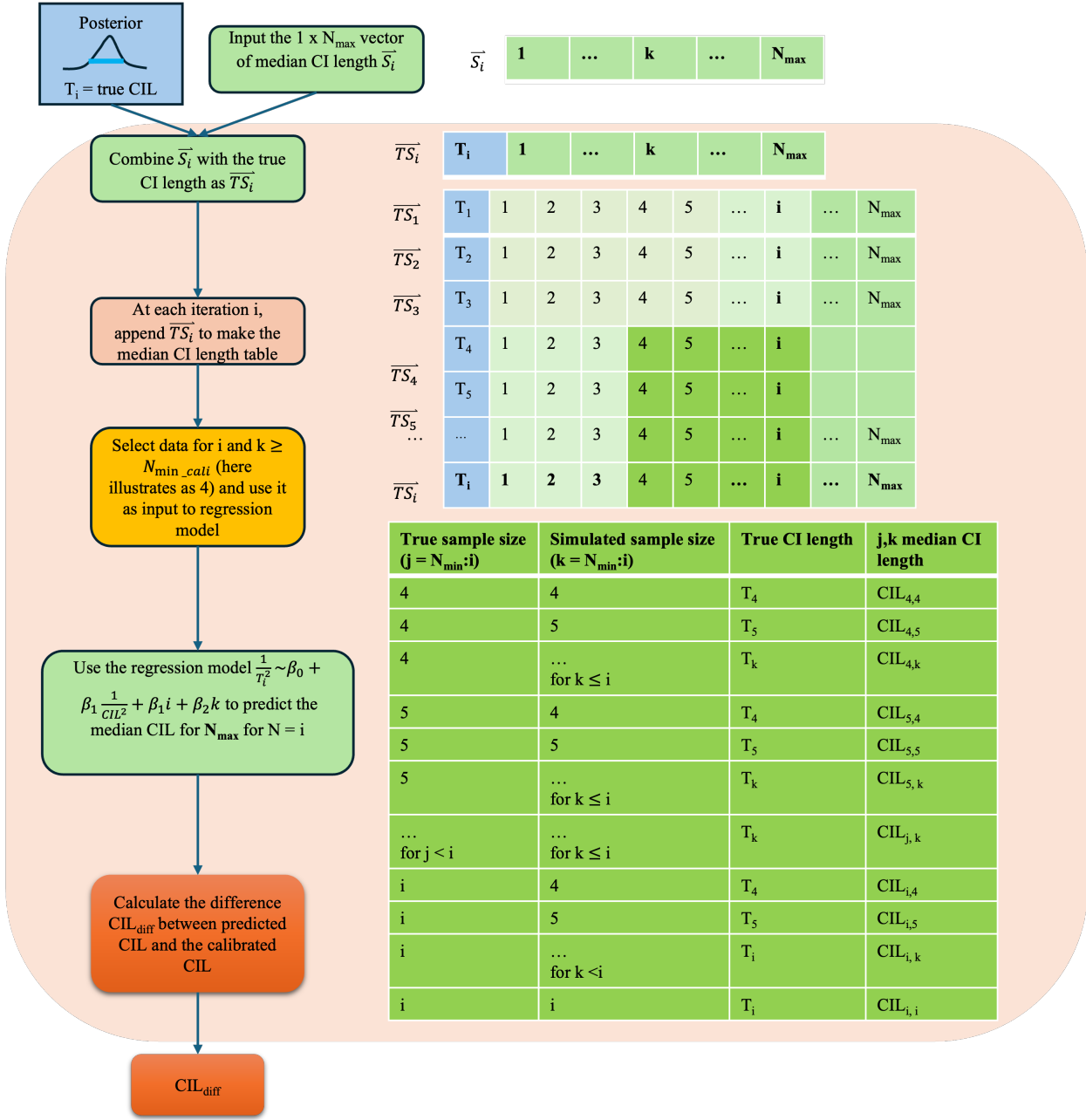

*Figure 4      Illustration of the calibration process. $\vec{S_i}$ is a vector of predicted median CIL and $T_i$ is the true data CIL. $\vec{TS_i}$ is the combined vector of $T_i$ and $\vec{S_i}$. With i increasing, $\vec{TS_i}$ is appended to previous $\vec{TS}$ in a table and this table can be used for regression modelling. $\beta_i$ are predictors.*

## 3   Statistical simulations
### 3.1   Simulation configuration
The goal of the statistical simulations is to explore the influence of different features of pBOS and compare pBOS to BOS and a frequentist sample size determination method with respect to precision targets. All simulations were run with a standard normal distribution N(0,1) as the experiment data model.

### 3.1.1 pBOS

To explore the performance and probability of pBOS, we ran statistical simulations manipulating four application-specific features: target CIL, TL, $N_{min}$, and the prior. This is followed by a description of how BOS and the fixed sample size (frequentist) approaches were configured.

#### 3.1.1.1 Target credible interval length (CIL)

Target CIL describes the required precision of the parameter of interest (for example the mean parameter in a normal distribution, as in this paper). Five target CILs were selected, based on the following percentiles of the CIL distribution of the true-data-based posterior CIL distribution at $N_{max}$: 5%, 25%, 50% 75%, and 95%. The range of percentiles represents statistical targets from "unlikely to meet the target" (5% CIL) to "likely to meet the target" (95% CIL), which covers most experimental situations.

#### 3.1.1.2 Tolerance level (TL)

The TL represents the probability of reaching the target in the future, based on predictions. The full tolerance range was explored from 0 to 1 in steps of 0.1. As previously mentioned, a tolerance of zero is the same as running BOS, because the experiment does not stop whether it is predicted to reach the target or not, and a tolerance of one is the same as always stopping at $N_{min}$. For the latter, as it is not possible to reach a 100% guarantee that the target can be reached, the decision algorithm will always suggest stopping when the sample size reaches $N_{min}$. A TL of 0.5 means that if the chance of reaching the target is lower than 50%, the experiment should stop.

#### 3.1.1.3 $N_{min}$

$N_{min}$ is the minimum sample size to allow for stopping the experiments, based on prediction. That is, if fewer than $N_{min}$ samples have been collected, the experiment can only be stopped with the traditional BOS (Part A in Figure 3); it will not stop even if predictions indicate that the target may not be reached (consequently such predictions do not need to be done until $N_{min}$). In this work $N_{max}$ was always set to 50 (see Section 3.1.2 for the motivation for this choice). Three values of $N_{min}$ were chosen: 10, 20, and 30—corresponding to 20%, 40%, and 60% (respectively) of the maximum data sample size ($N_{max}$)—which we argue represents a reasonable range for $N_{min}$.

#### 3.1.1.4 Prior distribution on mean and variance

Finally, five different priors (e.g., five sets of normal-gamma distributions) were explored. (See Figure 5 for an illustration.) A *central informative* prior exemplifies the case where high-quality, accurate information is available before the start for both mean and variance, which inherently helps reach a precision goal more quickly. "Central informative" here means that the mean and variance of the prior are the same as (or close to) the true mean and true variance, and there is relatively high trust in the prior (exemplified by a small variance). Two *weakly informative priors* were also explored. One has an accurate mean, but a larger variance than the central informative prior. The other, which has an incorrect mean (i.e., different from the true mean) and also a larger variance than the central informative prior, represents one of the most common situations when using Bayesian statistics (i.e., you typically cannot count on choosing a prior mean that is exactly correct, so it is usually a good idea to choose a weakly informative prior; Lemoine, 2019). The fourth prior is an *offset informative* prior: its narrow variance means that the prior information is believed accurate, but the mean is incorrect. The fifth prior, an almost completely *flat* prior, is essentially a Bayesian version of frequentist statistics— no prior information is used.

To simplify the process of exploring pBOS, we used conjugate priors, which allow the direct computation of posteriors, rather than using Markov Chain Monte Carlo to estimate the computation. Specifically, we used a normal data model and a normal-gamma prior with unknown mean and unknown variance. The prior model is shown in Equation (3). The data follow a normal distribution with $\mu$ as mean and $\frac{1}{\phi}$ as variance. $\phi$ follows a gamma distribution with the scaling parameter $v_0$, and $\mu$ is a normal distribution dependent on $\phi$ with the scaling parameter $n_0$. The two scaling parameters define the sample size for the prior. That is $n_0$ defines the sample size for mean parameter and $v_0$ defines the sample size for variance parameter. The sample size represents how much one trusts the prior. For example, when the sample size for the prior is 10 and the experiment has collected 10 samples, the prior and the data are equally weighted. $\mu_0$ is the prior mean and $\varphi_0$ is the prior variance.

$$\mu, \phi \sim NormGamma(\mu_0, n_0, \varphi_0, v_0) \tag{3}$$

$$\phi \sim Gamma(\frac{v_0}{2}, \frac{2}{v_0 \, \varphi_0}) \tag{4}$$

$$\mu|\phi \sim Norm(\mu_0, \frac{1}{\phi \, n_0}) \tag{5}$$

The choice of prior is an important component of Bayesian modeling and should be based on previous knowledge and literature. In this study, we conducted a sensitivity analysis by using the five different priors described above, in order to demonstrate the performance of pBOS given priors with different means and variances. This analysis is intended to help potential pBOS users decide if pBOS is an appropriate method for their application—and further, help users understand what prior to choose and how important the knowledge about the prior mean and variance is. The specific parameter values of the five priors used in the sensitivity analysis are shown in Table 1; their prior distributions are shown in Figure 5.

*Table 1        Parameter values of the five different kinds of priors*

|  | $\mu_0$ | $\varphi_0$ | $n_0$ | $v_0$ |
|---|---|---|---|---|
| Central informative prior | 0 | 1 | 10 | 10 |
| Central weakly informative prior | 0 | 10 | 5 | 1 |
| Offset weakly informative prior | 1 | 10 | 5 | 1 |
| Offset informative prior | 5 | 1 | 10 | 10 |
| Flat prior | 0 | 20 | 1 | 1 |

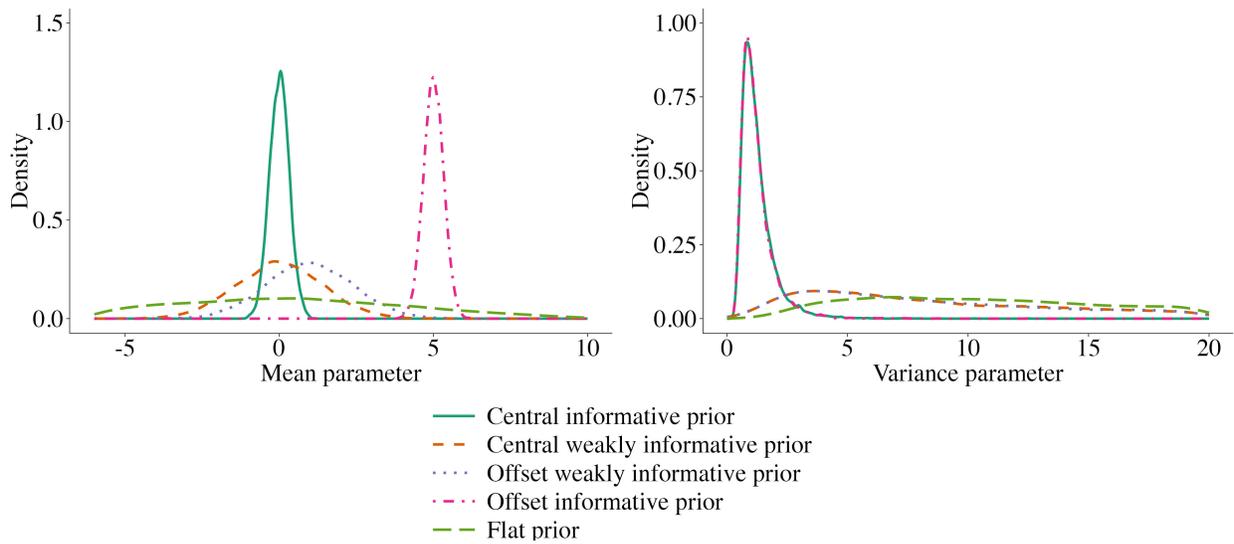

*Figure 5        Illustration of five priors' density distribution of mean and variance parameters.*

### 3.1.2 Sample size determination methods for comparison

The frequentist precision-based sample size determination method can be used to estimate the sample size needed to reach the target using equation (1). We assume the maximum available resource $N_{max}$ is 50. The CIL for pBOS with $N_{max}$ and five priors ranges from 0.4 to 0.7, so the confidence interval E is also chosen to go from 0.4 to 0.7. $Z$ is calculated as 1.96 with 95% confidence, and $\sigma$ is chosen between 1 and 3 (again, a range similar to the pBOS prior $\sigma$ range, which can be computed by taking the square root of the prior variance parameter $\varphi_0$ in Table 1). When E is set to 0.4, $\sigma$ is set to 1; when E is set to 0.7, $\sigma$ is set to 3. These two extremes produce an estimated required sample size between 24 and 71. In practice, the available resources limit sample size as well. For this paper we chose an $N_{max}$ of 50, which is within the calculated range; therefore, 50 was chosen both as the fixed sample size for the frequentist method and the $N_{max}$ in the simulations.

BOS was not implemented separately from pBOS, as it is just an extreme condition of pBOS. Recall that BOS is equivalent to pBOS when the rehearsal simulations for future predictions (see part B in Figure 3) always estimate that the CIL target can be reached within the TL—and therefore, no early stopping is initiated. That is, as mentioned in Section 3.1.1.2, in practice you can get BOS by setting the TL of pBOS to 0.

### 3.1.3 Performance evaluation

To assess the performance of the methods, we needed two components: the ground truth and the decision state. Here, the ground truth is whether the target is actually met or not in the course of an individual experiment; the truth can only be known after the data collection has been completed (i.e., $N_{max}$ has been reached). There are also two possible decision states for each experiment: to continue the experiment until the target is reached, or to stop the experiment early (when it is foreseen that the target cannot be reached). Consequently, there are four possible results, as shown in Table 2. This table is used as the basis for the receiver operating characteristic curve (ROC; de Hond et al., 2022) and the area under the ROC curve (AUC), which together serve as a means to evaluate the stopping-decision performance of pBOS across different feature settings. Huang & Ling (2005) have shown theoretically and empirically that AUC is a better criterion for consistency and discrimination than other criteria such as accuracy. In this work, AUC is calculated by summing the area under the ROC curve using the formula for the area of a trapezoid.

*Table 2       Four possible results for ground truth and decisions for each experiment.*

|  |  | Ground truth | |
|---|---|---|---|
|  |  | The goal can be met | The goal cannot be met |
| Decision | Continue experiment | True positive | False positive |
|  | Stop experiment early | False negative | True negative |

In addition to using ROC and AUC, we created a specific cost-benefit metric for assessing sample size determination methods, since experiments are often expensive and resources limited. The cost benefit per trial (where a trial is the collection of one sample) is calculated by dividing the benefit of reaching the target by the number of collected samples (the cost). To compare methods, one method is considered the norm (BOS) and each other method's cost benefit per trial is compared to this norm as a ratio. We treat the cost of each trial as C and the benefit of the experiment reaching its goal as B. The cost benefit per trial is thus $\frac{B}{C}$. The numbers of experiments that reach the target with pBOS and BOS are $n_{pBOS\_b}$ and $n_{BOS\_b}$, respectively, while the total number of trials for pBOS and BOS are $n_{pBOS\_c}$ and $n_{BOS\_c}$, respectively. The relative cost benefit for pBOS compared to BOS is shown in Equation (6). Because BOS is the norm, the cost-benefit proportion ratio is calculated by calculating the cost benefit of pBOS and fixed sample size minus the cost benefit of BOS first, and then dividing by the cost benefit of BOS.

$$R_{cb} = \frac{\frac{n_{pBOS\_b}B}{n_{pBOS\_c}C} - \frac{n_{BOS\_b}B}{n_{BOS\_c}C}}{\frac{n_{BOS\_b}B}{n_{BOS\_c}C}} = \frac{\frac{n_{pBOS\_b}}{n_{pBOS\_c}} - \frac{n_{BOS\_b}}{n_{BOS\_c}}}{\frac{n_{BOS\_b}}{n_{BOS\_c}}} \qquad (6)$$

As the benefit of reaching the target and the cost for each trial are the same across different methods, the relative cost-benefit performance metric is independent of the actual benefit and actual cost for one trial. This cost-benefit ratio shows how much better (if ratio is above zero) or worse (if the ratio is below zero) pBOS is compared to BOS. The ratio can also be calculated for fixed sample-size methods.

### 3.2 Results
#### 3.2.1 Calibration regression performance

Our results show that without the regression model's calibration, the pBOS prediction of the future CIL may substantially deviate from the ground truth CIL. Figure 6 shows that the median of the predicted future CIL is estimated to be higher than the ground truth CIL for the flat prior and smaller for the central informative prior. The CIL estimate for the flat prior at sample size 20 (with $N_{max}$ = 50) is 33.8% higher than the true CIL for $N_{max}$ 50, while the calibrated CIL is only 7.8% higher. As Figure 6 shows, as more data are collected and the calibration is updated, the error becomes smaller. At sample size 30, the predicted CIL is 21.5% higher than the true CIL, while the calibrated CIL is 4.6% higher. Unlike the flat prior, the informative prior overestimates the precision (smaller predicted CIL than the true CIL): the variance is smaller than the actual experiment's data model variance. However, in both cases the regression model substantially reduces the bias. Note that the overestimation case is likely to be rare in practice, whereas underestimating the precision (resulting in overly pessimistic power estimates) is likely to be more common. The median $R^2$ for the regression model ranges from 0.63 to 0.99 for the five priors across all combinations of the CIL, TL, and $N_{min}$.

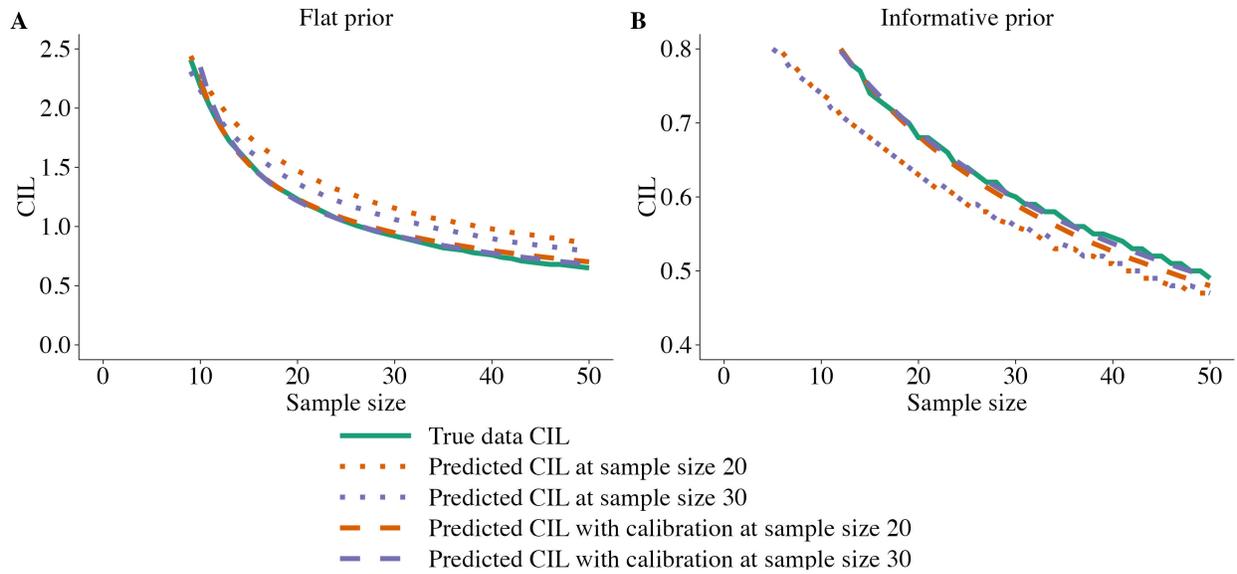

*Figure 6. CIL predictions for a flat prior (left panel) and a central informative prior (right panel). Predictions were made when 20 samples were collected (red) and when 30 samples were collected (blue), with calibration (dashed lines) and without (dotted lines). The dotted/dashed lines represent the predicted CILs for 200 simulations of each sample for each sample size (x-axis). Note that the informative prior variance is set to 0.5, which is smaller than the data variance of 1. The reason for choosing 0.5 was to illustrate the case of the overestimation of precision.*

### 3.2.2  Evaluation of the pBOS decision-making performance

The pBOS's performance is evaluated by a) AUC and b) the comparison to the performance of the BOS and the frequentist (fixed) sample size methods, using the proposed cost-benefit metric.

#### 3.2.2.1  ROC and AUC

Figure 7 shows an example of an ROC plot for the offset weakly informative prior under different CIL percentile criteria with $N_{min}$ of 10, 20, and 30. Table 3 shows the AUC values for the five priors under the same criteria with $N_{min}$ of 10. The range of the AUC for most priors is 0.79 to 0.98, except that the AUC for the offset informative prior has a range of 0.60 to 0.69. For the same CIL percentile condition, the central informative prior shows a larger AUC than the central weakly informative prior, which, in turn, shows a larger AUC than the offset weakly informative prior. These three priors all show larger AUCs than the flat prior. As expected, the offset informative prior shows the smallest AUC.

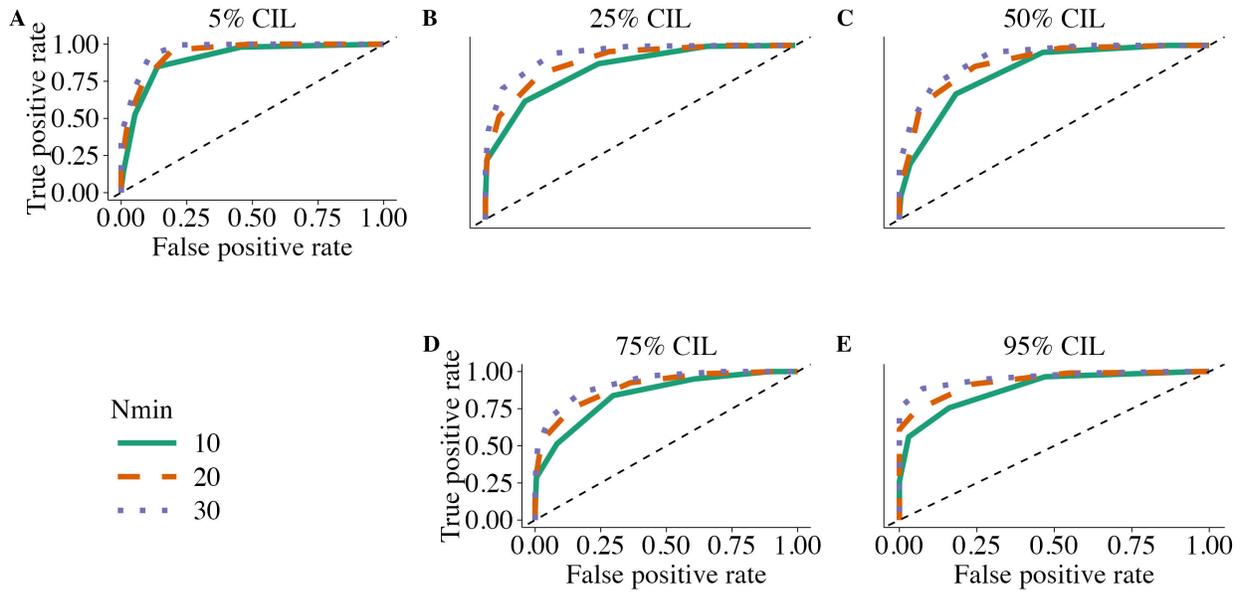

*Figure 7 Example of ROC plots for offset weakly informative prior based on 500 simulations.*

*Table 3 AUCs under different CIL percentiles across the five priors for $N_{min} = 10$.*

| CIL percentile | Central informative prior | Central weakly informative prior | Offset weakly informative prior | Offset informative prior | Flat prior |
|---|---|---|---|---|---|
| 0.05 | 0.98 | 0.94 | 0.90 | 0.68 | 0.85 |
| 0.25 | 0.95 | 0.88 | 0.87 | 0.60 | 0.79 |
| 0.5 | 0.94 | 0.87 | 0.85 | 0.68 | 0.81 |
| 0.75 | 0.94 | 0.87 | 0.84 | 0.69 | 0.78 |
| 0.95 | 0.96 | 0.88 | 0.89 | 0.67 | 0.81 |

#### 3.2.2.2 Cost benefit per trial

Two cost-benefit results are described. Figure 8 shows how the cost-benefit results vary across pBOS, BOS, and a fixed sample-size method as a function of the CIL target percentile and TL. Figure 9 shows the impact of the five assessed priors on the cost-benefit results.

As shown in Figure 8, the smaller the CIL threshold percentile, the more difficult it is to reach the target CIL at $N_{max}$, which means that pBOS will stop the experiment more often. That is, pBOS is better than BOS for low-CIL targets, while for high-CIL targets BOS is better than pBOS. This result indicates that pBOS has the better cost-benefit performance for experiments with high-precision statistical targets (since they are difficult to reach). When less precision is needed, BOS shows substantial advantages over both pBOS and fixed sample size.

With respect to the TL, our results show that pBOS has benefits over BOS from a TL of just above 0 to a maximum of 0.7 (for $N_{min}=10$ and a CIL target of 5%). The maximum cost-benefit increase ratio occurs at a TL of between 0.25 and 0.5, depending on the CIL target and the $N_{min}$.

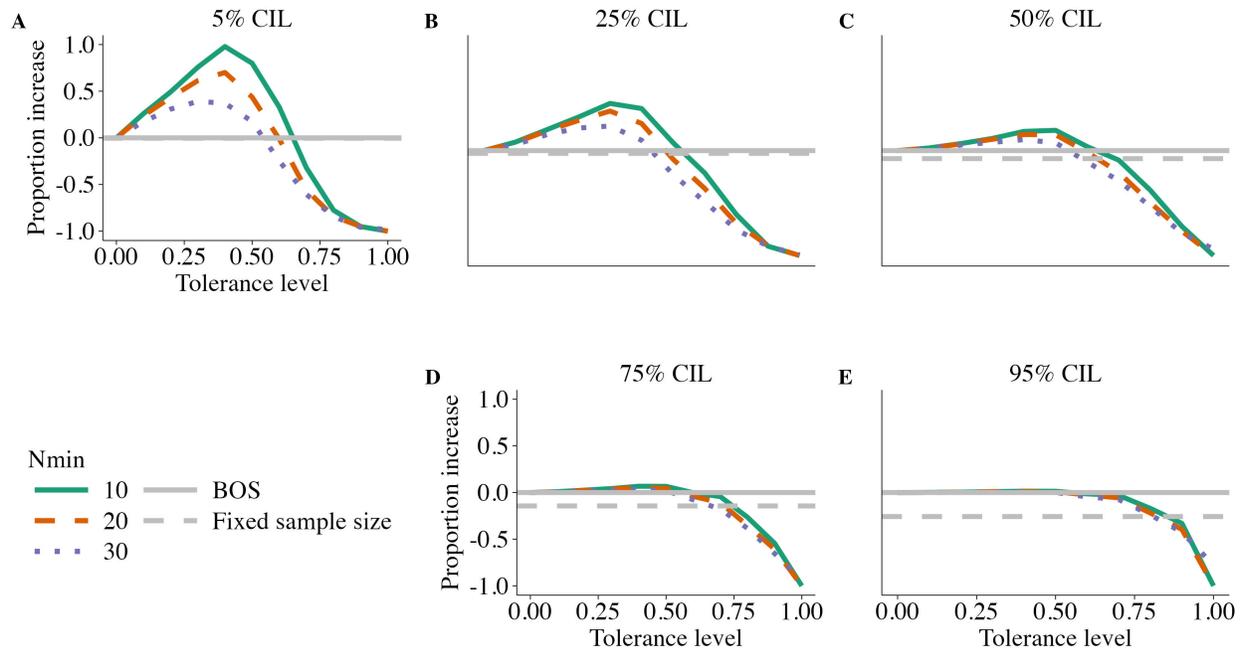

*Figure 8 The cost-benefit ratio relative to BOS for pBOS and fixed sample size and BOS for five target CILs (5% to 95%) as a function of the selected tolerance level (TL) and three choices of $N_{min}$. The y-axis is the proportional increase in cost benefit of the two other methods over BOS. At y=0, pBOS, fixed sample size, and BOS have the same cost benefit. For positive values, pBOS and fixed sample size are better than BOS, and for negative values, BOS is better. For pBOS and BOS the central informative prior is used (see Table 1).*

Figure 9 shows the cost-benefit performance of pBOS compared to BOS as a function of CIL and TL at $N_{min}$ = 10 for the different priors. Our results show that, as expected, the cost-benefit performance with the central informative prior is better for pBOS than for BOS; the weakly informative priors are also better to a lesser degree. The offset weakly informative prior shows a better cost benefit than the offset informative prior and the flat prior. For comparison, the solid green line in Figure 9 represents the central informative prior, shown in Figure 8.

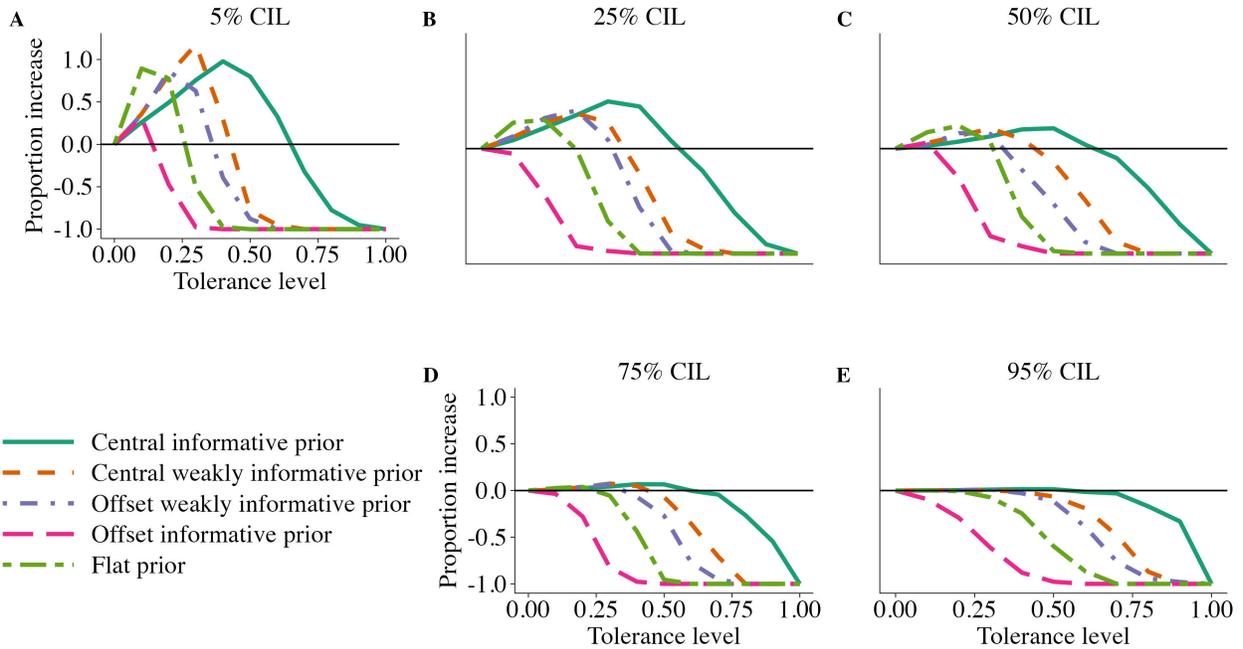

*Figure 9 A comparison of the cost benefits for pBOS across the five priors, for five CIL thresholds and as a function of tolerance level (TL; x-axis). See a description of how to read the graphs in Figure 8.*

## 4 Application in the traffic safety domain

As described in the introduction, we chose to apply pBOS to an experiment in traffic safety human factors to illustrate its use in practice. In this application, we seek to design a human-machine interface for an FCW system that minimizes the variability in response across drivers. Restated in the terms of this paper, the goal is to estimate the mean response time with a certain precision level. The advantage of this method becomes clear when experimental resources are limited; if the design under test is not promising, resources can be reallocated to testing a different design or to other development needs. This section describes the data, implementation choices, and results of the traffic safety application.

### 4.1 Data

An experiment was performed in a previous study that aimed to assess drivers' responses to an FCW (Puente Guillen & Gohl, 2019). As the number of samples collected in the original study was not large enough to demonstrate all aspects of pBOS, we did not use the actual data, but instead used their mean and variance to create a hypothetical "true" data-generation model. That data-generation model was then used to generate experiment samples for our demonstration. The following is a short overview of the original study.

The original study aimed to quantify how drivers react to an FCW and to investigate drivers' acceptance and trust in FCWs *(Puente Guillen & Gohl, 2019)*. The FCW was based on drivers' comfort zone boundaries. The data collected during the experiment include time-series recordings of drivers' driving behavior such as brake-pedal force, vehicle deceleration, and vehicle speed. The timing of the FCW trigger was also recorded. Based on those data, the time from the FCW trigger until the initiation of braking (based on brake-pedal force) was extracted *($T_{react}$)*. Figure 10 shows the drivers' reaction time's histogram plot.

As reaction times are often modeled as log normal distributions (Green, 2000), we also did that here. To be consistent in modeling using conjugate priors, the logs were first taken of the original data, and the mean and variance were then calculated based on the logs. We consequently used the mean and variance as a normally distributed data-generation model N(-0.054, $0.415^2$), assuming it to be the underlying data-generation model.

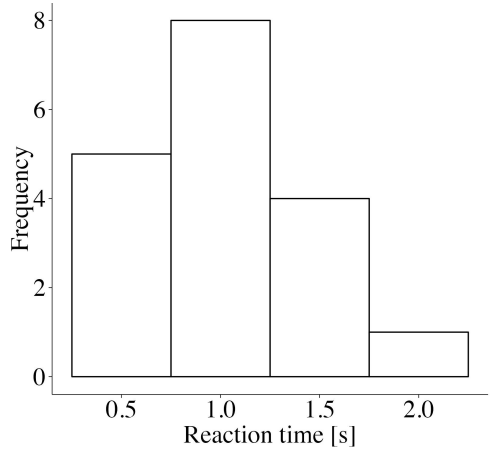

*Figure 10 Histogram plot of the drivers' reaction time to the FCW.*

## 4.2 Choosing pBOS parameters

The statistical simulations in Section 3 investigates the performance of pBOS for a variety of values of the four pBOS features (target CIL, TL, $N_{min}$, and the prior). However, only one configuration (set of feature values) is chosen for a given experiment. The choice of configuration should be based on an understanding of how each feature influences pBOS' performance for the specific application under study. Further, for each feature, any additional suggestions for determining its value are described below, along with the specific choices made for the FCW application. Additional guidelines for feature-value selection can be found in the Discussion.

*Target CIL.* For this FCW experiment demonstration, we selected a target CIL of 0.3 s. The specific choice should be based on experts' suggestions about the needed precision of the reaction time mean. We considered that +-0.15 s a reasonable target CIL (this is similar to the idea of Bayesian region of practical equivalence as a precision goal; Kruschke, 2014). According to literature (Green, 2000), the mean reaction time to unexpected events in traffic is 1.5 s. The values of the estimated lower and upper CIL bounds were calculated by subtracting the log of the expected mean (1.5s) +- half the CIL. Thus, the target CIL in the log unit was chosen to be the result of subtracting log(1.35) from log(1.65), resulting in a $CIL_{thres}$ of 0.30.

*$N_{min}$.* The value needs to be large enough to support the initial rehearsal simulations and calibration, but small enough to require as few trials as possible. $N_{min}$ was chosen to be 10 samples.

*TL.* Since pBOS is most efficient when the TL is set to a value between 0.2 to 0.6, we chose 0.4, as it is the peak value for the relative cost benefit for pBOS over BOS (see Figure 9).

*Prior.* As mentioned in the previous section, a conjugate normal prior with unknown mean and variance was chosen. The reaction time mean was set to 1.5 s and the variance to 0.5, in accordance with Green (2000). However, as estimates of the reaction time mean and variance vary in the literature and this particular FCW has not been assessed before, a weakly informative prior was chosen. Thus, $n_0$ was set to 5 and $v_0$ to 1.

## 4.3 Analysis and results

After values for the four features were chosen, the experiment and analysis could start. Recall that at this point it was unknown whether the CIL target was likely to be reached within $N_{max}$ samples. The first check was performed when $N_{min}$ samples had been collected. At this point, a relatively low probability (less than 50%) would indicate that the target CIL is a strict target, and therefore probably difficult to reach. The estimated cost benefit for this case is shown in the first three plots in Figure 8; pBOS could be a better choice than BOS. However, if the target CIL percentile is estimated to be larger than 50%, BOS is probably the best choice, as illustrated in Figure 8.

To assess the performance of the proposed pBOS, we simulated many potential FCW experiments. We first created 100,000 groups of data from the data-generation model N(-0.054, 0.415$^2$) to explore the randomness of the FCW experiment. The ground truth (how likely it was that the experiment would reach the target) was established by combining the data with the selected (weakly informative) prior. We calculated the CIL at $N_{max}$ (50); the probability of reaching the CIL target for these 100,000 groups of data turned out to be 8%. This value suggests that the target could be reached for 8% of the generated data sets. This probability is low—much lower than 50%. To avoid low accuracy in estimation due to low occurrence of "could reach targets", we randomly selected an equal number (100) of "could reach the target" and "could not reach the target" data sets from the original groups of data. We then started the stopping decision simulations as if we were running an actual experiment, having chosen the BOS or pBOS methods (based on the paragraph above to guide the stopping decision during the experiment).

For each of these 200 groups of data, we simulated the stopping decision as if we were running an actual experiment, having chosen the BOS or pBOS methods. We then weighted the decision accuracy results by multiplying the ground truth's weight of 0.08 to the "could reach the target" groups and 0.92 to the "could not reach the target" groups. The cost benefit per trial of this practical combination of pBOS and BOS was 17% higher than when only BOS was used. This result suggests that including pBOS has the potential to save approximately 17% of the allocated resources, making them available for alternative uses.

# 5 Discussion

This work proposes and demonstrates a method that can be used to halt an experiment either when sufficient data has been collected, or when it is predicted that a specific statistical target (e.g., a threshold on the CIL) is unlikely to be reached, given the experimental constraints. The latter decision can potentially save time and resources. The method, predictive Bayesian optional stopping (pBOS), is a combination of the traditional Bayesian optional stopping (BOS) method and the use of rehearsal simulations as described in Kruschke's (2014) Bayesian posterior-based sample size determination approach. The proposed method is agnostic to the choice of stopping criteria, but we used a precision-target criterion based on CIL in our comparison of pBOS to both the traditional BOS with CIL as the target and to a fixed sample-size method (equivalent to using a precision target-based frequentist sample size determination approach).

## 5.1 General results

The performance of the proposed pBOS method was evaluated in two ways. First, we treated the pBOS decision to stop an experiment, based on the prediction "will reach target" or "will not reach target", as an algorithm, and then assessed its performance against the ground truth, using the AUC metric. Across the five priors tested, the median AUC ranged from 0.62 to 0.97 (0.74 to 0.97 when the offset weakly informative prior, which may be unusual in practice, was excluded). The ROC literature labels this range as at least moderate, fair, or acceptable (de Hond et al., 2022), which indicates that the performance of the proposed method is relatively effective and robust. Second, we developed a method to assess the cost benefit of experiment sample size determination methods that reflects the use case of limited experimental resources. The performance of pBOS was further compared with BOS and the use of a fixed sample size with different target CILs and TLs. In the following we discuss our manipulation of the four pBOS features for which decisions need to be made when setting up pBOS, and how these features influence pBOS performance.

The evaluated features were a) CIL target, b) TL, c) $N_{min}$, and d) the choice of prior. The results indicate that pBOS outperforms BOS under a subset of the combinations of feature choices. Specifically, pBOS is more cost-effective than BOS when a) the CIL target is difficult to reach and b) there is enough information available from previous work to select at least a weakly informative prior (see Figures 8 and 9). In this scenario, pBOS can accurately predict when the experiment will *fail* to reach the statistical target. The experimenter can stop the experiment early, with the option of either revising it (e.g., modifying some experimental condition) or redirecting resources elsewhere. Our results indicate that pBOS generally has a positive cost benefit over BOS if the underlying probability of reaching the target CIL is under 50%. In practice, this can be estimated using the probability of reaching the target at $N_{max}$ when $N_{min}$ samples have been collected. On the other hand, if the probability of reaching the target is greater than 50%, the traditional BOS method is preferred.

In practice, the CIL target should be chosen based on the application, typically based on expert knowledge, the literature, and/or previous experiments. The recommendation for the choice of $N_{min}$ is based on the statistical simulation results, which show that smaller values have higher cost benefits— even though, as shown in Figure 7, the ROC results show better performance with larger $N_{min}$. The primary advantage of pBOS is that it provides an opportunity to stop the experiment early so that resources can be redirected more constructively. The earlier "could not reach target" experiments are stopped, the more resources can be saved and the better the cost benefit.

The selection of the prior should also be based on expert knowledge. It is well known that an informative prior can make the estimation more efficient (as also demonstrated in Figure 9). However, and also as expected, a wrong informative prior performance is even worse than a flat prior. Given our results, we recommend the use of a weakly informative prior, as it includes part of the benefits of an informative prior and avoids the risk of choosing a wrong informative prior.

After setting up the CIL target, $N_{min}$, and prior, the next step is to check whether the experimental context meets the pBOS-beneficial condition: the estimated probability of reaching the target CIL is below 50% with $N_{min}$ collected data. The remaining two features of pBOS can, to some extent, be selected to maximize the efficiency of the experiment. As shown in Figure 8, pBOS is efficient when the target is hard to reach (target CIL is narrow) and TL is relatively small. The recommended TL for the pBOS-beneficial conditions (less than 50% CIL target) is between 0.2 and 0.6. In general, a more inclusive (i.e., smaller) TL improves the cost benefit. Somewhat counter-intuitively, the more inclusive TL, combined with a hard-to-reach CIL goal, optimizes the benefit of being able to redirect resources under some circumstances (based on the hard-to-reach goal) but not too often quit too early (based on the inclusive TL), which would mean rarely meeting the target. Based on the pBOS statistical simulation results, the recommended range of TL is between 0 and 0.6. In this range, pBOS shows no worse cost benefit than BOS for most cases. If such constraints allow a TL in the proposed range to be chosen, pBOS efficiency will be improved.

To assess the robustness of pBOS with respect to different prior parameters, we conducted a sensitivity analysis. We compared pBOS for five different types of priors: a central informative prior, a central weakly informative prior, an offset weakly informative prior, an offset informative prior and a flat prior. Our results (see Figure 9) show trends similar to those of the cost-benefit analysis: there is an increase and then a decrease as TL ranges from 0 to 1 across the different priors, for the same CIL target. As expected, the central informative prior performs the best and the offset informative prior the worst, while the weakly informative priors show better performance than the offset informative prior and the flat prior. Further, the central weakly informative prior is better than the offset weakly informative prior. These findings demonstrate that, as with all Bayesian methods, pBOS is most beneficial when one has some reasonably accurate prior information. If such information is not available, a weakly informative prior is the recommended choice. However, our results show that pBOS is not a method you should typically choose if you have a flat prior.

Note that users of the pBOS must recognize that if they decide to stop early because it is unlikely they can collect enough data to achieve the statistical target, there will not be sufficient information to achieve the goal of the experiment. However, not reaching the statistical target does not mean the data are useless; they might still be useful in a traditional Bayesian analysis (Schönbrodt & Wagenmakers, 2018) for decision-making (depending on the application).

## 5.2 Precision misestimation and calibration of the target

In pBOS, the decision whether to stop an experiment is based on the analysis of predicted future samples. As previously described, similar to Kruschke (2014), the predicted future samples are simulated from the data model whose parameters are taken from the posterior distribution based on the prior and the collected data. The posterior is assumed to be the underlying data-generation model, which means that the generated data include all "potential experiments". In contrast, when a real experiment is run, the generated data are only from one specific data-generation instantiation (i.e., the equivalent of one set of parameters from the posterior-based data-generation model). Consequently, the generated future data in pBOS are likely to have a larger variance (CIL) and lower precision than in a real experiment. This situation usually leads to a misestimate of the variance in pBOS' rehearsal-simulation generated-data posterior distribution. Note that for general applications there is no explicit analytical solution for calculating this bias. However, practically, it can always be established using Markov Chain Monte Carlo estimations.

The calibration step in pBOS addresses this precision misestimation (typically an underestimation). We performed the calibration for the normal gamma distribution prior by analytically calculating the CIL for data generated with rehearsal simulations and compared it with the CIL for (simulated) single-experiment instantiations (assuming the full experiment is played out). The result is typically a precision underestimation, but when the prior is informative, the result might be an overestimation instead.

To compensate for the precision misestimation, we propose a calibration. Empirical results of the regression model (Figure 6) show how it compensates for misestimation. The median of $R^2$ for the regression model ranges from 0.63 to 0.99, which suggests a robust fit for the regression model based on the data. The identification of the underestimation of the precision estimated by pBOS and the corresponding regression-based compensation, based on rehearsal simulations, is another contribution of this paper. That is, the parameters that could influence the prediction model are chosen as predictors in the regression model. Consequently, the choice of the predictors should depend on the application. It is also worth noting that the rehearsal simulation approach proposed by Kruschke (2014) could also suffer from this misestimation, although that is not mentioned in that work, and we have not found it discussed elsewhere. In Appendix A we demonstrate this misestimation analytically for the case when the prior is a conjugate normal gamma distribution.

## 5.3 Benefit evaluation methods

Enabling resource-conditioned early stopping can liberate previously allocated resources, which can be applied instead to adjust the current experiments or revise the research question—or they can be used in other experiments. In this way, pBOS maximizes productivity with the available resources, producing better cost-benefits.

In our application, the stopping target includes CIL and TL. As the TL relates to the probability of achieving the target with limited resources, the choice of TL is linked to cost-benefit. Most of the experiment approaches have a hypothesis-testing perspective, which focuses on the decision accuracy; on the other hand, another approach to optional stopping has a cost-efficiency perspective, based on predicted costs and quality (Wagner et al., 2023). The CIL and TL targets can be combined with the cost-benefit decision-making when the benefits are known. If the real-world benefits of achieving the target are known and can be quantified, then the potential real-world benefit for a specific TL (probability of achieving the target) can be estimated by multiplying the real-world benefit by the TL. Here, a real-world benefit may be a substantial increase in system performance or a cost reduction – quantified in some way. Also, the cost can be estimated by the experiment trial costs in the same way as in the FCW demonstration. The decision of whether to continue the experiments can be made based on the estimated cost benefit by subtracting the estimated cost from the estimated benefit.

## 5.4 Stopping criteria – choice and implications

The choice of stopping criteria is crucial in optional stopping. Although many papers with the term BOS in the title or keywords apply optional stopping using Bayes factors (Schmalz et al., 2021), BOS (and pBOS) is actually a much more extensive methodology that can be applied using other stopping criteria, such as precision targets, as in this work. The debate on the appropriateness of using significance testing-based stopping criteria (e.g., Bayes factors) for BOS is still ongoing. Yu et al. (2014) demonstrated that Bayesian optional stopping does not eliminate the biases that exist in optional stopping used in frequentist tests, and de Heide & Grünwald (2021) argue that BOS can break down when default or pragmatic priors are used for the parameters of interest. Rouder (2014) used simulations to demonstrate that Bayesian statistics, including posterior odds and Bayes factors, are *not* affected by the stopping rule; the author also emphasized the proper interpretation of Bayesian quantities as subjective belief and the difficulty of using frequentist intuition to interpret Bayesian statistics. Rather than focusing on a specific stopping criterion, the current paper proposes a generic method with the flexibility to adopt different criteria, depending on the application.

## 5.5 The pBOS application to the traffic safety domain

Due to the frequentist replication crisis and lesser concerns about the computation efforts of Bayesian methods (Wagenmakers et al., 2012), there is growing interest in applying Bayesian statistics in general, and BOS in particular. Examples include online experiments (Wan et al., 2023), clinical trials (Pourmohamad & Wang, 2023), psychological research (Vasishth et al., 2023), management research (Andraszewicz et al., 2015), and survey studies (Wagner et al., 2023). However, we have not been able to find any study that uses BOS in the traffic safety or human-machine-interface (HMI) design application domain, even though it is common to conduct experiments both in traffic safety research and in system development.

There is one sub-area of the traffic safety domain where BOS and pBOS may be particularly useful: human factors engineering (HFE). In HFE, a wide range of experiments may be conducted, often including drivers as study participants. For example, the impact of a specific safety system or HMI design might be explored. Experiments assessing driver responses often have high variability, due to the heterogeneity of human participants (Liu et al., 2023).

Applying the proposed method in these experiments could let them be stopped early—as soon as the precision (CIL) of the parameter of interest is either satisfactory or unlikely to be reached with the maximum experiment resources available. The parameter of interest could be reaction time, as in the FCW application, or off-road glance distributions (i.e., how drivers look off the forward roadway when, e.g., engaging in some tasks; Donmez et al., 2010). Another HFE use-case where a CIL target-based pBOS may be beneficial is when trying to minimize the variability in eyes-off-road glances (as an aspect of HMI design). It has been shown that different tasks (and HMIs) result in different variances in driver off-road glance distributions, which affect the safety of the system (Reimer et al., 2012; Ulahannan et al., 2022). If the variance of a task or HMI can be minimized (within and across drivers), it becomes easier to predict what drivers will do, and design distraction countermeasures (e.g., driver monitoring systems) accordingly.

Using pBOS, one could conduct experiments and test HMIs by quantifying driver off-road glance distributions with a pre-defined target precision, to identify an HMI design with accepted high precision of the off-road glance distribution and then design distraction countermeasures for that HMI design. The experiment could start with a relatively small (but realistic) CIL target; if it does not seem possible to reach the target, the experiment is stopped, the HMI design can be tweaked (as can the prior, since there is now more information), and another set of participants can be run. Decisions can be made iteratively using the pBOS flow. A combination of BOS and pBOS in an experiment means that resources could be reallocated if it proves to have a low chance of success. These methods could, in principle, be combined with Response Surface Methods (RSM; Bezerra et al., 2008) and other space-searching methods that iteratively look for combinations of conditions that have the best desired outcome (based on a target similar to that used for pBOS). The goals of pBOS (and BOS) are not the same as RSM, but they are compatible—RSM could be used to determine where to look next, and pBOS could be used to stop early when a given HMI design is unlikely to reach the goal.

As previously noted, when an experiment is run and stopped by pBOS, it is not necessarily a waste of resources; as in all applications of Bayesian statistics, the data collected can be useful even if you stop early and "significant" (or in the case of using a CIL target, high-precision) results are not obtained. For example, even if the target is not reached, it may be possible to use the information about the CIL to improve the design of the FCW. For the off-road glance application, the CIL can be used as a prior in future studies (i.e., help setting more correct priors), as well as for design improvement.

## 5.6 Limitations and future work

The benefits of pBOS are greatest in experiments that are expensive and do not have large amounts of data, since stopping early avoids unnecessary waste. On the other hand, if the experiment is not costly in terms of time or money, the benefits of pBOS may may be trivial.

Bayesian approaches such as BOS and pBOS are typically much more computationally demanding (especially when conjugate models cannot be used) than frequentist methods. However, the rapidly increasing availability of computing power means that this requirement is not likely to be a major hurdle for the future use of these approaches.

The demonstration of pBOS was performed only for a precision-based target, while many experiments aim to determine effect size or significance. Limiting the scope in this way entailed a more straightforward and likely accepted methodology, which seemed appropriate for an initial assessment of pBOS. Nonetheless, a lack of a significance-testing target use-case application is a limitation. This could be explored in future work.

The proposed multiple regression model for bias compensation is modeled from an engineer's perspective, not a statistician's. That is, the aim is to get the method 'good enough' rather than precise and based on statistical theory. The dependent and independent variables are chosen practically to calibrate the predictions. The calibration can be improved and future work should investigate this calibration from the perspectives of statistical theory.

In this work, pBOS only uses the collected data in the first step to generate parameter posterior distribution for future data-generation models. Whether it is appropriate to include the collected data in subsequent steps as part of the Bayesian analysis is not investigated. Future work includes investigating the influence of combining the experimentally collected data with the simulated future data in the rehearsal simulations on the FCW application (which includes only the simulated future data).

# 6 Conclusions

The pBOS method proposed in this paper, a combination of the traditional BOS with Kruschke's Bayesian sample size determination method, can, under certain conditions, reduce the resources needed in experiments. Our results confirm previous work demonstrating that BOS is more cost-efficient than frequentist sample-size determination. In addition, our results indicate that it is beneficial to use pBOS when there is less than 50% probability of reaching the target; the cost-benefit advantage over BOS can be as much as 118%. However, when the target is not difficult to reach (reachable with more than 50% probability), it is typically better to use the traditional BOS. Whether the target is difficult to meet or not can be determined by an analysis of the initial data collected, and the choice between pBOS and BOS can be made based on the estimation. The cost-benefit performance of pBOS is influenced by the simulation configuration—specifically, by the choices of four features. To assess the impact of different feature values, we also provide a sensitivity analysis and guidelines for choosing each feature.

## Acknowledgement

The authors thank the European Commission for funding this work through the SHAPE-IT project under the European Union's Horizon 2020 research and innovation programme (Marie Skłodowska-Curie grant agreement 860410). The authors would like to thank András Bálint from Volvo Car Corporation (at the time of contributing he was a Chalmers employee) for his valuable inputs during the initial concept formalization phase of this work. We also want to thank the PROSPECT project for conducting the study we used as a basis for our demonstration of the method. The authors also want to thank Kristina Mayberry (Mayberry Academic Services) for her language review.

## Declaration of generative AI in scientific writing

During the preparation of this work the authors used a generative large language model on some small parts of the manuscript in order to refine the text and improve clarity. After using this tool, the authors reviewed and edited the content as needed and take full responsibility for the content of the publication.

# Appendix A

This appendix demonstrates the use of a normal-gamma conjugate prior for normally distributed data. The calculation proceeds from the data and prior distribution to their posterior distribution and corresponding variance of the mean parameter. Subsequently, based on this posterior distribution, the predicted data with the same prior's posterior distribution and its corresponding variance of the mean parameter are computed. We demonstrate that the variance of the mean parameter's posterior distribution based on the predicted data can differ from the variance of the mean parameter's posterior distribution based on the collected data. The relationship between these two variances depends on the prior and the data collected.

*Assume that iid samples of data $X = x_1, x_2, x_3, \ldots, x_n$ follow a normal distribution with an unknown mean $\mu$ and unknown variance $\varphi$ where $X|\mu, \varphi \sim N(\mu, \varphi)$, and where the prior for $\mu, \varphi$ follows a normal gamma distribution shown in Equation A. 1 with $\phi = 1/\varphi$.*

$$\mu, \varphi \sim NormGamma(\mu_0, n_0, \varphi_0, v_0) \qquad A.\ 1$$

$$\phi \sim Gamma\left(\frac{v_0}{2}, \frac{2}{v_0 \varphi_0}\right) \qquad A.\ 2$$

$$\mu|\phi \sim N\left(\mu_0, \frac{1}{n_0 \phi}\right) \qquad A.\ 3$$

*By combining the prior for $\mu, \varphi$ with the collected data (sample size=n), the posterior distribution can be computed and is shown in Equation A. 4.*

$$\mu, \varphi \sim NormGamma(\mu_1, n_1, \varphi_1, v_1) \qquad A.\ 4$$

*The parameters inside Equation A. 4 are as below where s is the sample data variance.*

$$\mu_1 = \frac{n_0 \mu_0 + n\bar{x}}{n_0 + n} \qquad A.\ 5$$

$$n_1 = n_0 + n \qquad A.\ 6$$

$$v_1 = v_0 + n \qquad A.\ 7$$

$$\varphi_1 = \frac{1}{v_1}\left[(n-1)s^2 + v_0\varphi_0 + \frac{nn_0}{n_1}(\bar{x} - \mu_0)^2\right] \qquad A.\ 8$$

*As the marginal distribution of $\mu$ follows a T distribution, the $1 - \alpha$ highest density interval for $\mu$ with collected data can be expressed as Equation A. 9.*

$$E(ci_1) = 2t_{cirt}\sqrt{\frac{E(\varphi_1)}{n_1}} = 2t_{cirt}\sqrt{\frac{\frac{1}{v_1}\left[(n-1)s^2 + v_0\varphi_0 + \frac{nn_0}{n_1}(\bar{x} - \mu_0)^2\right]}{n_0 + n}} \qquad A.\ 9$$

*The future possible data samples are generated by the data model whose parameters come from the posterior distribution based on the collected data ($X = x_1, x_2, x_3, \ldots, x_n$). If the independent simulated data $Y = y_1, y_2, y_3, \ldots, y_{n'}$ follows a normal distribution with an unknown mean $\mu$ and variance $\varphi$ where $Y|\mu, \varphi \sim N(\mu, \varphi)$, and the prior for $\mu, \varphi$ is the same as in Equation A. 1, then the simulated data posterior distribution can be expressed as Equation A. 10.*

$$\mu, \varphi \sim NormGamma(\mu_2, n_2, \varphi_2, v_2) \qquad A.\ 10$$

The parameters inside Equation A. 10 are as below where $n'$ is the simulated data size and $s_2$ is the sample variance of the simulated data.

$$\mu_2 = \frac{n_0\mu_0 + n'\bar{y}}{n_0 + n'} \qquad A.\ 11$$

$$n_2 = n_0 + n' \qquad A.\ 12$$

$$v_2 = v_0 + n' \qquad A.\ 13$$

$$\varphi_2 = \frac{1}{v_2}[(n'-1)s_2^2 + v_0\varphi_0 + \frac{n'n_0}{n_2}(\bar{y} - \mu_0)^2] \qquad A.\ 14$$

As in Equation A. 9, the $1 - \alpha$ highest density interval for $\mu$ with simulated data can be expressed as Equation A. 15.

$$E(ci_2) = 2t_{cirt}\sqrt{\frac{E(\varphi_2)}{n_2}} = 2t_{cirt}\sqrt{\frac{\frac{1}{v_2}[(n'-1)s_2^2 + v_0\varphi_0 + \frac{n'n_0}{n_2}(\bar{y}-\mu_0)^2]}{n_0 + n'}} \qquad A.\ 15$$

As the parameters for the simulated data come from the posterior distribution based on the prior and the collected data, $Y|\mu_s, \phi_s \sim N(\mu_s, 1/\phi_s)$ and $\mu_s, \phi_s$ can be expressed in Equation A. 16 and A. 17 with $\phi_s = \frac{1}{\varphi_s}$.

$$\phi_s \sim Gamma\left(\frac{v_1}{2}, \frac{2}{v_1\varphi_1}\right) \qquad A.\ 16$$

$$\mu_s|\phi_s \sim N(\mu_1, \frac{1}{n_1\phi_s}) \qquad A.\ 17$$

As $Y$ follows a normal distribution, a statistic can be created as in Equation A. 18 that is distributed as chi-square with $df = n' - 1$.

$$\frac{(n'-1)s_2^2}{\frac{1}{\phi_s}} \sim \chi^2_{n'-1} \qquad A.\ 18$$

As the mean of $\chi^2_{n'-1}$ is $n' - 1$, the expected value of $s_2^2$ can be expressed as in Equation A. 19.

$$E(s_2^2) = E\left(\frac{1}{\phi_s}\right) = \frac{1}{E(\phi_s)} \qquad A.\ 19$$

As shown in Equation A. 16, $\phi_s$ follows a gamma distribution, so the expected value of $\varphi_s$ can be expressed as in Equation A. 20.

$$E(s_2^2) = \frac{1}{E(\phi_s)} = \frac{1}{\frac{v_1}{2}\frac{2}{v_1\varphi_1}} = E(\varphi_1) \qquad A.\ 20$$

By combining the Equation A. 19 and Equation A. 20,

$$E(s_2{}^2) = \frac{1}{E(\phi_s)} = E(\varphi_1) \qquad A.\ 21$$

*As the simulated data Y follows a normal distribution, $Y|\mu_s, \varphi_s \sim N(\mu_s, \varphi_s)$, $\bar{y}$ also follows a normal distribution as in Equation A. 22.*

$$\bar{y} \sim N\left(\mu_s, \frac{1}{\phi_s n'}\right) \qquad A.\ 22$$

*Therefore,*

$$E[(\bar{y} - \mu_0)^2] = Var(\bar{y}) + [E(\bar{y} - \mu_0)]^2 = \frac{1}{E(\phi_s) n'} + (\mu_s - \mu_0)^2 \qquad A.\ 23$$

*By combining the Equation A. 5, A. 21 and A. 23 into Equation A. 15,*

$$E(\varphi_2) = \frac{1}{n' + v_0}\left[\left(n' - 1 + \frac{n_0}{n' + n_0}\right)E(\varphi_1) + v_0 \varphi_0 + \frac{n' n_0 n^2}{n' + n_0}\left(\frac{\bar{x} - \mu_0}{n + n_0}\right)^2\right] \qquad A.\ 24$$

*When the simulated data sample size is the same as the collected data sample size, $n' = n$, therefore,*

$$E(\varphi_2) = \frac{[(n-1)(n+n_0) + n_0]}{(n+v_0)(n+n_0)} E(\varphi_1) + \frac{v_0 \varphi_0}{n+v_0} + \frac{n_0 n^3}{(n+n_0)(n+v_0)}\left(\frac{\bar{x} - \mu_0}{n + n_0}\right)^2] \qquad A.\ 25$$

$\varphi_1$ is the posterior variance with the collected data ($X = x_1, x_2, x_3, \ldots, x_n$) and prior and $\varphi_2$ is the posterior variance with the predicted data ($Y = y_1, y_2, y_3, \ldots, y_{n'}$) and the same prior. When $n' = n$, the sample size of the collected data and predicted data is the same. As shown in the Equation A. 25, $\varphi_2$ is related to $\varphi_1$ but the relative value is influenced by the prior and the collected data.

There are three terms in the Equation A. 25. $\frac{[(n-1)(n+n_0)+n_0]}{(n+v_0)(n+n_0)}$ is the coefficient for $E(\varphi_1)$, and $\frac{v_0 \varphi_0}{n+v_0} + \frac{n_0 n^3}{(n+n_0)(n+v_0)}\left(\frac{\bar{x}-\mu_0}{n+n_0}\right)^2$] are the additional two terms. As $n_0$ $v_0$ $\varphi_0$ and $n$ are all positive, then: the first and second term must be positive and the third term is non-negative.

Therefore, there are three possible outcomes:

- When the coefficient $\frac{[(n-1)(n+n_0)+n_0]}{(n+v_0)(n+n_0)} < 1$ and the additional terms are not large enough to make up the difference, $E(\varphi_2) < E(\varphi_1)$.
- When the coefficient $\frac{[(n-1)(n+n_0)+n_0]}{(n+v_0)(n+n_0)} < 1$ and the additional terms are just enough to make up the difference, $E(\varphi_2) = E(\varphi_1)$.
- When the coefficient $\frac{[(n-1)(n+n_0)+n_0]}{(n+v_0)(n+n_0)} < 1$ and the additional terms are large enough to make up the difference or when the coefficient $\frac{[(n-1)(n+n_0)+n_0]}{(n+v_0)(n+n_0)} \geq 1$, $E(\varphi_2) > E(\varphi_1)$.

Consider the limit as $n_0$ approaches infinity in the coefficient term.

$$\lim_{n_0 \to \infty}[(n-1)(n+n_0) + n_0] = \lim_{n_0 \to \infty} n_0(n+1) \qquad A.\ 26$$

$$\lim_{n_0 \to \infty}(n+v_0)(n+n_0) = \lim_{n_0 \to \infty} n_0(n+v_0) \qquad A.\ 27$$

$$\lim_{n_0 \to \infty} \frac{[(n-1)(n+n_0)+n_0]}{(n+v_0)(n+n_0)} = \frac{\lim_{n_0 \to \infty}[(n-1)(n+n_0)+n_0]}{\lim_{n_0 \to \infty}(n+v_0)(n+n_0)} = \frac{n+1}{n+v_0} \qquad A.\ 28$$

As $v_0$ is the sample size for $\varphi_0$, $v_0 \geq 1$. So when $n_0$ approaches infinity, the coefficient term $\frac{[(n-1)(n+n_0)+n_0]}{(n+v_0)(n+n_0)}$ decreases and $\frac{[(n-1)(n+n_0)+n_0]}{(n+v_0)(n+n_0)} \leq 1$. $v_0$ appears only in the denominator in the coefficient term. So when $v_0$ increases, the coefficient term $\frac{[(n-1)(n+n_0)+n_0]}{(n+v_0)(n+n_0)}$ decreases. When the coefficient term decreases, $E(\varphi_2)$ becomes smaller and more dependent on the additional terms.

As $\varphi_0$ only appears in the numerator in $\frac{v_0 \varphi_0}{n+v_0}$. When $\varphi_0$ decreases, $\frac{v_0 \varphi_0}{n+v_0}$ also decreases and $E(\varphi_2)$ decreases. When $v_0$ approaches infinity, the value of $\frac{v_0 \varphi_0}{n+v_0}$ is $\varphi_0$, shown in Equation A. 29.

$$\lim_{v_0 \to \infty} \frac{v_0 \varphi_0}{n+v_0} = \varphi_0 \lim_{v_0 \to \infty} \frac{v_0}{n+v_0} = \varphi_0 \qquad A.\ 29$$

For the other additional term $\frac{n_0 n^3}{(n+n_0)(n+v_0)}\left(\frac{\bar{x}-\mu_0}{n+n_0}\right)^2$, the smaller the difference between $\bar{x}$ and $\mu_0$, the smaller it is. In addition, $v_0$ only appears in its denominator, so the bigger the $v_0$, the smaller this term. When $n_0$ increases, the denominator increases cubically with $n_0$ and the numerator increases linearly with $n_0$. As the denominator increases faster than the numerator when $n_0$ increases, the term $\frac{n_0 n^3}{(n+n_0)(n+v_0)}\left(\frac{\bar{x}-\mu_0}{n+n_0}\right)^2$ decreases when $n_0$ increases as well.

In summary, a larger $n_0$, $v_0$ and a smaller $\varphi_0$ and difference between $\bar{x}$ and $\mu_0$ can lead to a smaller $E(\varphi_2)$ and possibly $E(\varphi_2) \leq E(\varphi_1)$. Otherwise, $E(\varphi_1)$ can be bigger than $E(\varphi_2)$.

As the $ci_1$ and $ci_2$ are calculated based on $\varphi_1$ and $\varphi_2$ respectively, the relative value of $ci_1$ and $ci_2$ are also influenced by the prior and the collected data.